\documentclass[runningheads]{svmult}

\usepackage{makeidx}   
\usepackage{graphicx}  
\usepackage{subeqnar}  
\usepackage{multicol}  
\usepackage{physprbb}  
\makeindex             

\newcommand{\lsim}{\rlap{\raise 2pt \hbox{$<$}}{\lower 2pt
\hbox{$\sim$}}\ }
\newcommand{\gsim}{\rlap{\raise 2pt \hbox{$>$}}{\lower 2pt
\hbox{$\sim$}}\ }
\newcommand{\dslash}{\rlap{/}{\partial}}
\newcommand{\Aslash}{\rlap{/}{A}}
\newcommand{\viola}{\rlap{/}}
\newcommand{\s}{{\rm s}}

\begin{document}
\title*{Neutrinos in physics and astrophysics}
\toctitle{Neutrinos in physics and astrophysics}
%
%
\titlerunning{Neutrinos in physics and astrophysics}
%
\author{Esteban Roulet}
\authorrunning{Esteban Roulet}
%
%
\institute{Departamento de F\'\i sica\\
Universidad Nacional de La Plata\\
CC67, 1900, La Plata, Argentina}

\maketitle              

\begin{abstract}
An elementary general overview of the neutrino physics and
astrophysics is given. We start by a historical account of the
development of our understanding of neutrinos and how they helped to
unravel the structure of the Standard Model. We discuss why it is so
important to establish if neutrinos are massive and we introduce the main
scenarios to provide them a mass. The present bounds and the positive
indications in favor of non-zero neutrino masses are discussed as well
as the major role they play in astrophysics and cosmology.

\end{abstract}

\section{The neutrino story:}

\subsection{The hypothetical particle:}

One may trace back the appearance of neutrinos in physics to the 
discovery of radioactivity by Becquerel one century ago. When the
energy of the electrons (beta rays) emitted in a radioactive decay
 was measured by Chadwick in 1914, it turned out to his surprise 
to be continuously distributed. 
This was not to be expected if the underlying process
in the beta decay was the transmutation of an element $X$ into 
another one $X'$ with the emission of an electron, i.e. $X\to X'+e¯$, 
since in that case the electron should be monochromatic. The situation 
was so puzzling that Bohr even suggested that the conservation
of energy may not hold in the weak decays. Another serious problem
with the `nuclear models' of the time was the belief that
nuclei consisted of protons and electrons, the only known particles 
by then. To explain the mass and the charge of a 
 nucleus it was then necessary that it 
had $A$ protons and $A-Z$ electrons in it. For instance,
a $^4$He nucleus would have
 4 protons and 2 electrons. Notice that this total of 
six fermions would make the $^4$He nucleus to be a boson, which 
is correct. However, the problem arouse when this theory was applied
for instance to $^{14}$N, since consisting of 14 protons and 7 electrons
would make it a fermion, but the measured angular momentum of the
nitrogen nucleus was $I=1$.

The solution to these two puzzles was suggested by Pauli only in 1930, 
in a famous letter to the `Radioactive Ladies and Gentlemen' 
gathered in a meeting in Tubingen, where he wrote: `I have hit 
upon a desperate remedy to save the exchange theorem of statistics 
and the law of conservation of energy. Namely, the possibility that 
there could exist in nuclei electrically neutral particles, that I wish to 
call neutrons, which have spin 1/2 ...'. These had to be not heavier
than electrons and interacting not more strongly than gamma rays.

With this new paradigm, the nitrogen nucleus became 
$^{14}$N$=14 p+7e+7`n$', which is a boson, 
and a beta decay now involved the emission of two particles
$X\to X'+e+`n$', and hence the electron spectrum was continuous. 
Notice that no particles were created in a weak
decay, both the electron and Pauli's neutron $`n$' were already 
present in the nucleus of the element $X$, and they just came 
out in the decay.
However, in 1932 Chadwick discovered the real `neutron', with a mass 
similar to that of the proton and being the missing building block
of the nuclei, so that a nitrogen nucleus finally became just
$^{14}$N$=7p+7n$, which also had  the correct bosonic statistics.

In order to account now for the beta spectrum of weak decays, 
Fermi called Pauli's hypotetised particle the neutrino 
(small neutron), $\nu$, 
and furthermore suggested that the fundamental process underlying
beta decay was $n\to p+e+\nu$. He wrote \cite{fe34} 
the basic interaction by analogy with the interaction known at the time,
the QED, i.e. as a vector$\times$vector current interaction:

$$H_F=G_F\int {\rm d}^3x[\bar\Psi_p\gamma_\mu\Psi_n]
[\bar\Psi_e\gamma^\mu\Psi_\nu]+ h.c. .$$
This interaction accounted for the continuous beta spectrum, and 
 from the measured shape at the endpoint Fermi concluded that
$m_\nu$ was consistent with zero and had to be small.
The Fermi coupling $G_F$ was estimated from the observed lifetimes 
of radioactive elements, and armed with this Hamiltonian 
Bethe and Peierls \cite{be34} decided to compute the cross section
for the inverse beta process, i.e. for $\bar\nu+p\to n+e^+$, which was
the relevant reaction to attempt the direct detection of a neutrino. 
The result, $\sigma=4(G_F^2/\pi)p_eE_e\simeq 2.3\times 10^{-44}$cm$^2
(p_eE_e/m_e^2)$ was so tiny that they concluded `... This meant that
one obviously would never be able to see a neutrino.'. For instance, 
if one computes the mean free path in water (with density 
$n\simeq 10^{23}/$cm$^3$) 
of a neutrino with energy $E_\nu=2.5$ MeV, typical of a weak decay, 
the result is $\lambda\equiv 1/n\sigma\simeq 2.5\times 10^{20}$ cm, 
which is $10^7$AU, i.e. comparable to the thickness of the 
Galactic disk.

It was only in 1958 that Reines and Cowan  were able to prove
that Bethe and Peierls had been too pessimistic, when they 
measured for the first time the interaction of a neutrino 
through the inverse beta process\cite{re59}. Their strategy was 
essentially that, if one needs $10^{20}$ cm of water to stop a neutrino, 
 having $10^{20}$ neutrinos a cm would be enough to stop one neutrino.
Since after the second war powerful reactors started to become available,
and taking into account that in every fission of an uranium
nucleus the neutron rich fragments beta decay producing typically
6 $\bar\nu$ and liberating $\sim 200$ MeV, it is easy to show that
the (isotropic) neutrino flux at a reactor is
$${d\Phi_{\nu}\over d\Omega}\simeq {2\times 10^{20}\over 4\pi}
\left( {{\rm Power\over GWatt}}\right){\bar\nu\over strad}.$$  
Hence, placing a few hundred liters of water near a reactor they
were able to see the production of positrons (through the two 
511 keV $\gamma$ produced in their annihilation with electrons) and 
neutrons (through the delayed $\gamma$ from the neutron capture in Cd), 
with a rate consistent with the expectations from the weak interactions
of the neutrinos.

\subsection{The vampire:}

Going back in time again to follow the evolution of the 
theory of weak interactions of neutrinos, in 1936 Gamow and Teller
\cite{ga36} noticed that the $V\times V$ Hamiltonian of Fermi was
probably too restrictive, and they suggested the generalization

$$H_{GT}=\sum_iG_i[\bar pO_in][\bar eO^i\nu]+h.c. ,$$
involving the operators $O_i=1,\ \gamma_\mu,\ \gamma_\mu\gamma_5,\ 
\gamma_5,\ \sigma_{\mu\nu}$, corresponding to scalar ($S$), 
vector ($V$), axial vector $(A)$, pseudoscalar ($P$) and tensor ($T$)
currents. However, since $A$ and $P$ only appeared here as $A\times A$ or 
$P\times P$, the interaction was parity conserving.
The situation became unpleasant, since now there were five different
coupling constants $G_i$ to fit with experiments, but however
this step was required since some observed 
nuclear transitions which were
forbidden for the Fermi interaction became now allowed with its
generalization (GT transitions). 

The story became  more involved when in 1956 Lee and Yang
suggested that parity could be violated in weak
interactions\cite{le56}.  This 
could explain why the particles theta and tau had exactly the
same mass and charge and only differed in that the first one 
was decaying to two pions while the second to three pions (e.g. to
states with different parity). The explanation to the 
puzzle was that the $\Theta$ and $\tau$  were 
just the same particle, now known as the charged kaon, but the 
(weak) interaction leading to its decays violated parity.

Parity violation was confirmed the same year by Wu \cite{wu57}, 
studying the direction of emission of the electrons emitted in
the beta decay of polarised $^{60}$Co. The decay rate is
 proportional to $1+\alpha \vec{P}\cdot \hat{p}_e$. 
Since the Co polarization vector $\vec P$ is an axial vector, while the
unit vector along the electron momentum $\hat{p}_e$ is a vector, their
scalar product is a pseudoscalar and hence a non--vanishing 
coefficient $\alpha$ would imply parity violation. The result was 
that electrons preferred to be emitted in the direction opposite 
to $\vec{P}$, and the measured 
value $\alpha\simeq -0.7$ had then profound implications for
the physics of weak interactions. 

The generalization by  Lee and Yang of the Gamow Teller Hamiltonian
was
$$H_{LY}=\sum_i[\bar pO_in][\bar eO^i(G_i+G_i'\gamma_5)\nu]+h.c. .$$
Now the presence of terms such as $V\times A$ or $P\times S$ allows
for parity violation, but clearly the situation became even more 
unpleasant since there are now 
10 couplings ($G_i$ and $G_i'$) to determine, so that some order was
really called for.

Soon the bright people in the field realized that there could be
a simple explanation of why parity was violated in weak interactions,
the only one involving neutrinos, and this had just to do with
the nature of the neutrinos. Lee and Yang, Landau and Salam
\cite{bright} realized that, if the neutrino was
massless, there was no need to have both neutrino chirality states 
in the theory, and hence the handedness of the neutrino could be the
origin for the parity violation. 
 To see this, consider the chiral projections of a fermion
$$\Psi_{L,R}\equiv {1\mp \gamma_5\over 2}\Psi.$$
We note that   
in the relativistic limit these two projections describe
left and right handed helicity states (where the helicity, i.e. the
spin projection in the direction of motion, is a constant of motion
for a free particle), but in general an helicity eigenstate is a
mixture of the two chiralities. For a massive particle, which
has to move 
 with a velocity smaller than the speed of light, it is always
possible to make a boost to a system where the helicity is reversed, 
and hence the helicity is clearly not a Lorentz invariant while the
chirality is (and hence has the desireable properties of a charge to
which a gauge boson can be coupled).
If we look now to the equation of motion for a Dirac particle
 as the one we are used to for the description of
a charged massive particle such as an electron ($(i\dslash-m)\Psi=0$), 
 in terms of the chiral projections this equation becomes
$$i\dslash\Psi_L=m\Psi_R$$
$$i\dslash\Psi_R=m\Psi_L$$
and hence clearly a mass term will mix the two chiralities.
However, from these equations we see that for $m=0$, as could be the
case for the neutrinos, the two equations are decoupled, and one 
could write a consistent theory using only one of 
the two chiralities (which in this case would coincide with the
helicity).  If the Lee Yang Hamiltonian were just to depend
on a single neutrino chirality, one would have then $G_i=\pm G_i'$ and 
parity violation would indeed be
maximal. This situation has been described by
saying that neutrinos are like vampires in Dracula's stories: 
if they were to look to
themselves into a mirror they would be 
unable to see their reflected images.

The actual helicity of the neutrino was measured
by Goldhaber et al. \cite{go58}.
The experiment consisted in observing the $K$-electron capture
in $^{152}$Eu ($J=0$) which produced  $^{152}$Sm$^*$ ($J=1$) 
 plus a neutrino. This excited nucleus then decayed into 
$^{152}$Sm ($J=0)+\gamma$. Hence the measurement of the polarization 
of the photon gave the required 
information on the helicity of the neutrino 
emitted initially. The conclusion was that `...Our results
seem compatible with ... 100\% negative helicity for the neutrinos', 
i.e. that the neutrinos are left handed particles.

This paved the road for the $V-A$ theory of weak interactions
advanced by Feynman and Gell Mann, and Marshak and 
Soudarshan \cite{vma}, which stated that weak interactions only
involved vector and axial vector currents, in the combination 
$V-A$ which only allows the coupling to left handed fields, i.e.
$$J_\mu=\bar e_L\gamma_\mu\nu_L+\bar n_L\gamma_\mu p_L$$
with $H=(G_F/\sqrt{2})J_\mu^\dagger J^\mu$.
This interaction also predicted the existence of purely leptonic weak 
charged currents, e.g. $\nu+e\to \nu+e$, to be experimentally 
observed much later\footnote{A curious fact was that the new theory
predicted a cross section for the inverse beta decay a factor of two
larger than the Bethe and Peierls original result, which 
had already been  confirmed in 1958 to the 5\% accuracy by 
Reines and Cowan. However, in a
new experiment in 1969, Reines and Cowan found 
 a new value consistent with the new
prediction, what shows that many times when the experiment agrees with
the theory of the moment the errors tend to be underestimated.}. 

The current
involving  nucleons is actually 
not exactly $\propto \gamma_\mu(1-\gamma_5)$
(only the interaction at the quark level has this form), but
is instead $\propto \gamma_\mu(g_V-g_A\gamma_5)$. The vector
coupling remains however
 unrenormalised ($g_V=1$) due to the so called conserved
vector current hypothesis (CVC), which states that the vector part of the
weak hadronic charged currents ($J_\mu^\pm\propto 
\bar \Psi\gamma_\mu \tau^\pm\Psi$, with $\tau^\pm$ the raising 
and lowering operators in the isospin space $\Psi^T=(p,n)$) together with
the isovector part of the electromagnetic current (i.e. the term 
proportional to $\tau_3$ in the decomposition $J^{em}_\mu\propto
\bar \Psi\gamma_\mu (1+\tau_3)\Psi$) form an isospin triplet
of conserved currents. On the other hand, the axial vector hadronic
current is not protected from strong interaction renormalization effects
and hence $g_A$ does not remain equal to unity. The measured value,
using for instance the lifetime of the neutron, is $g_A=1.26$, so that
at the nucleonic level the charged current 
weak interactions are actually ``$V-1.26A$''.

With the present understanding of weak interactions, we know that
the clever idea  to explain parity 
violation as due to the non-existence of one of the neutrino
chiralities (the right handed one) was completely wrong, although
it lead to major advances in the theory and ultimately 
to the correct interaction. Today we understand that the parity violation 
is a property of the gauge boson (the $W$) responsible for the gauge 
interaction, which  couples only to the left handed fields, 
and not due to the
absence of right handed fields. For instance, in the quark sector both
left and right chiralities exists, but parity is violated because
the right handed fields are singlets for the weak charged currents.

\subsection{The trilogy:}

In 1947 the muon was discovered in cosmic rays by Anderson and Neddermeyer.
This particle was just a heavier copy of the electron, and as was
suggested by Pontecorvo, it also had weak interactions $\mu +p\to n+\nu$ 
with the same universal strength $G_F$. Hincks, Pontecorvo and Steinberger
showed that the muon was decaying to three particles, $\mu\to e\nu\nu$, 
and the question arose whether the two emitted neutrinos were 
similar or not. It was then shown by Feinberg \cite{fe58} that, assuming
the two particles were of the same kind, weak interactions couldn't be
mediated by gauge bosons (an hypothesis suggested in 1938 by Klein).
The reasoning was that if the two neutrinos were equal, it would be
possible to join the two neutrino lines and attach a photon to the
virtual charged gauge boson ($W$) or to the external legs, 
so as to generate a diagram for
the radiative decay $\mu\to e\gamma$. The resulting branching ratio
would be larger than $10^{-5}$ and was 
hence already excluded at that time.
This was probably the first use of `rare decays' to constrain 
the properties of new particles.

The correct explanation for the absence of the radiative decay
was put forward by Lee and Yang, 
who suggested that the two neutrinos
emitted in the muon decay had different flavour, 
i.e. $\mu\to e+\nu_e+\nu_\mu$, and hence it was not possible 
to join the two neutrino lines to draw the  radiative decay diagram. 
This was confirmed at Brookhaven 
in the first accelerator neutrino experiment\cite{da62}. 
They used an almost
pure $\bar\nu_\mu$ beam,
something which can be obtained from charged 
pion decays, since the $V-A$ theory
implies that $\Gamma(\pi\to \ell+\bar\nu_\ell)\propto m_\ell^2$, i.e. 
this process requires a chirality flip in the final lepton line which
strongly suppresses the decays $\pi\to e+\bar\nu_e$.
Putting a detector in front of this beam they were able to observe
the process $\bar\nu+p\to n+\mu^+$, but no production of positrons, 
what proved that the neutrinos produced in a weak decay in association
with a muon were not the same as those produced in a beta
decay (in association with an electron).
Notice that although the neutrino fluxes are much smaller at 
accelerators than at reactors, their higher energies make their detection
feasible due to the larger cross sections ($\sigma\propto E^2$ 
for $E\ll m_p$, and $\sigma\propto E$ for $E\gsim m_p$).

In 1975 the $\tau$hird charged lepton was discovered by Perl at
SLAC, and being just
a heavier copy of the electron and the muon, it was concluded that
a third neutrino flavour had also to exist. Although the direct 
detection through e.g. $\bar\nu_\tau+p\to n+\tau^+$ has not yet been 
possible, due to the difficulty of producing a $\nu_\tau$ beam and of
detecting the very short $\tau$ track, there is
little doubt about its existence, and we furthermore know today that
the number of light weakly interacting neutrinos is precisely three
(see below), so that the proliferation of neutrino species seems
to be now under control.

\subsection{The gauge theory:}

As was just mentioned, Klein had suggested that the short range
charged current weak interaction could be due to the exchange of
a heavy charged vector boson, the $W^\pm$. 
This boson exchange would look at
small momentum transfers ($Q^2\ll M_W^2$) as the non renormalisable 
four fermion interactions discussed before. If the gauge interaction
is described by the Lagrangian ${\cal L}=-(g/\sqrt{2})J_\mu W^\mu+h.c.$,
from the low energy limit one can identify the Fermi coupling as
$G_F=\sqrt{2}g^2/(8M_W^2)$.
In the sixties, Glashow, Salam and Weinberg showed that it was
possible to write down  a unified description of electromagnetic
and weak interactions with a gauge theory based in the group $SU(2)_L\times
U(1)_Y$ (weak isospin $\times$ hypercharge, with the electric charge
being $Q=T_3+Y$), 
which was spontaneously broken at the weak scale down 
to the electromagnetic $U(1)_{em}$. This (nowadays standard) model 
involves the three gauge bosons in the adjoint of $SU(2)$, $V_i^\mu$ (with
$i=1,2,3$), and the hypercharge gauge field $B^\mu$, so that the starting 
Lagrangian is 
$${\cal L}=-g\sum_{i=1}^3J^i_\mu V^\mu_i-g'J^Y_\mu B^\mu + h.c. ,$$
with  $J^i_\mu\equiv \sum_a \bar \Psi_{aL}\gamma_\mu (\tau_i/2)\Psi_{aL}$.
The left handed leptonic and quark
isospin doublets are $\Psi^T=({\nu_e}_L,e_L)$ and 
($u_L, d_L)$ for the first generation (and similar ones for the other two
heavier generations) and the right handed fields are SU(2) singlets. 
The hypercharge current is obtained by summing 
over both left and right handed fermion chiralities and is
$J^Y_\mu\equiv \sum_a Y_a\bar \Psi_{a}\gamma_\mu \Psi_{a}$.

After the electroweak breaking one can identify the
weak charged currents with $J^\pm=J^1\pm iJ^2$, which couple to the
$W$ boson $W^\pm=(V^1\mp iV^2)/\sqrt{2}$, and the two neutral vector
bosons $V^3$ and $B$ will now combine through a rotation by the 
weak mixing angle $\theta_W$ (with tg$\theta_W=g'/g$), to give
\begin{eqnarray}
\pmatrix{ A_\mu \cr Z_\mu}=\pmatrix{ {\rm c}\theta_W & \s\theta_W\cr
-\s\theta_W & {\rm c}\theta_W} \pmatrix{ B_\mu\cr V^3_\mu}.
\end{eqnarray}
We see that the broken theory has now, besides the massless photon field
$A_\mu$,  an additional neutral vector boson, the heavy $Z_\mu$, 
whose mass 
turns out to be related to the $W$ boson mass through 
$s^2\theta_W=1-(M_W^2/M_Z^2)$. The electromagnetic and neutral weak
currents are given by 
$$J_\mu^{em}=J^Y_\mu+J^3_\mu$$
$$J^0_\mu=J^3_\mu-s^2\theta_W J^{em}_\mu,$$
with the electromagnetic coupling being $e=g\ \s\theta_W$.

The great success of this model came in 1973 with the experimental
observation of the weak neutral currents using muon neutrino beams
at CERN (Gargamelle) and Fermilab, using the elastic 
process $\nu_\mu e\to \nu_\mu e$. The semileptonic processes
$\nu N\to \nu X$ were also studied and the comparison of neutral
and charged current rates provided a measure of the weak mixing angle.
From the theoretical side t'Hooft proved the renormalisability of the
theory, so that the computation of radiative corrections became 
also meaningful.

\begin{figure}[t]
\begin{center}
\includegraphics[width=.5\textwidth]{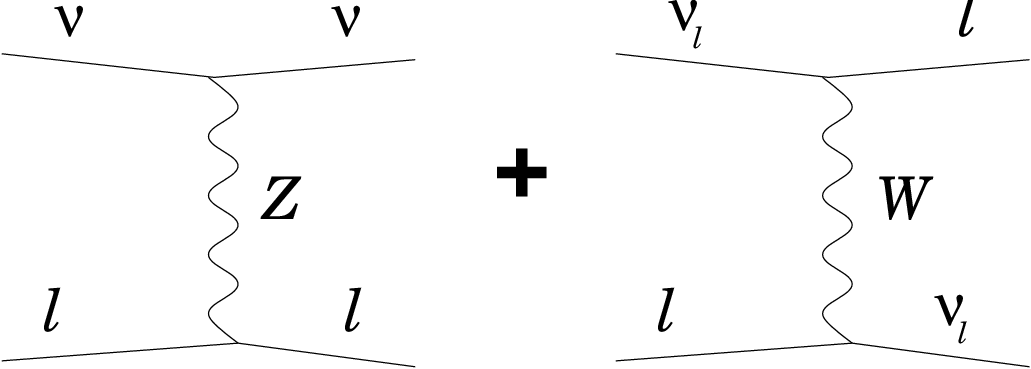}
\end{center}
\caption[]{Neutral and charged current contributions to neutrino 
lepton scattering.}
\label{eps1}
\end{figure}

The Hamiltonian for the leptonic weak interactions 
$\nu_\ell+\ell'\to\nu_\ell+\ell'$ can be obtained, 
using the Standard Model just presented, from the two diagrams 
in figure 1. In the low energy limit ($Q^2\ll M_W^2,\ M_Z^2$), 
it is just given by
$$H_{\nu_\ell\ell'}={G_F\over \sqrt{2}}[\bar\nu_\ell \gamma_\mu
(1-\gamma_5)\nu_\ell][\bar\ell'\gamma^\mu(c_LP_L+c_RP_R)\ell']$$
where the left and right couplings are $c_L=\delta_{\ell\ell'}+
s^2\theta_W-0.5$ and $c_R=s^2\theta_W$. The $\delta_{\ell\ell'}$ term 
in $c_L$ is due to the charged current diagram, which clearly only
appears when $\ell=\ell'$. On the other hand, one sees that
due to the $B$ component in the $Z$ boson, the weak neutral currents
also  couple to the charged lepton right handed chiralities 
(i.e. $c_R\neq 0$).
This interaction leads to the cross section (for $E_\nu\gg m_{\ell'}$)
$$\sigma(\nu+\ell\to \nu+\ell)={2G_F^2\over \pi}m_\ell 
E_\nu\left[c_L^2+{c_R^2 \over 3}\right],$$
and a similar expression with $c_L\leftrightarrow c_R$ for antineutrinos.
Hence, we have the following relations for the neutrino
 elastic scatterings off electrons 
$$\sigma(\nu_e e)\simeq 9\times 10^{-44}{\rm cm}^2\left(
{E_\nu\over 10\ {\rm MeV}}\right)\simeq 2.5\sigma(\bar\nu_e e)
\simeq 6\sigma(\nu_{\mu,\tau}e)\simeq 7\sigma(\bar\nu_{\mu,\tau}e).$$
Regarding the angular distribution of the electron momentum with 
respect to the incident neutrino direction, in the center
of mass system of the process $d\sigma(\nu_e e)/d\cos\theta\simeq
1+0.1 [(1+\cos \theta)/2]^2$, and it is hence almost isotropic. However, 
due to the boost to the laboratory system, there will be a 
significant correlation between the neutrino and electron momenta
for $E_\nu\gg $MeV, and this actually allows 
to do astronomy with neutrinos.
For instance, water cherenkov detectors such as Superkamiokande detect
solar neutrinos using this process, and have been able to reconstruct
a picture of the Sun with neutrinos. It will turn also to be relevant
for the study of neutrino oscillations that these kind of detectors 
are six times more sensitive to electron type neutrinos than to the
other two neutrino flavours.

Considering now the neutrino nucleon interactions, one has at low
energies (1~MeV$<E_\nu<50$~MeV)
$$\sigma(\nu_en\to p e)\simeq \sigma(\bar\nu_ep\to n e^+)\simeq 
{G_F^2\over \pi}{\rm c}^2\theta_C(g_V^2+3g_A^2)E_\nu^2,$$
where we have now
introduced the Cabibbo mixing angle $\theta_C$ which relates,
 if we ignore the third family, 
the quark flavour eigenstates $q^0$ to the mass eigenstates $q$, 
i.e. $d^0={\rm c}\theta_C d+\s\theta_Cs$ and $s^0=-\s\theta_C
d+{\rm c}\theta_Cs$ (choosing a flavour basis so that  the 
up type quark flavour and mass eigenstates coincide).

\begin{figure}[t]
\begin{center}
\includegraphics[width=.4\textwidth,angle=-90]{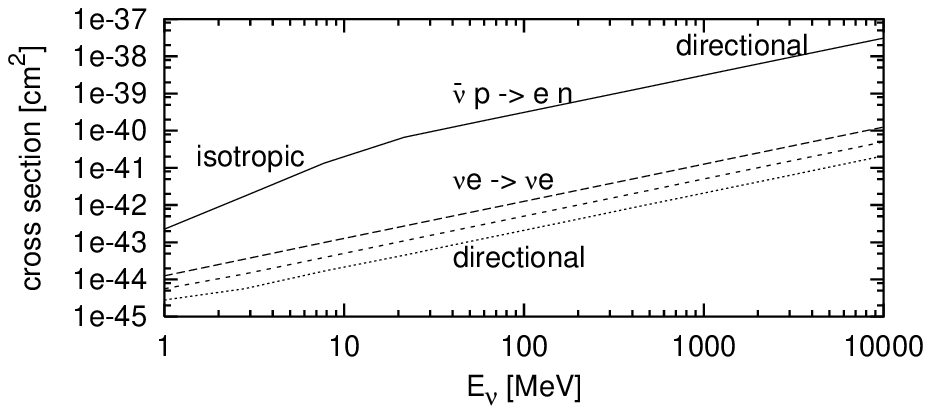}
\end{center}
\caption[]{Neutrino nucleon and neutrino lepton cross sections (the three 
lines correspond, from top to bottom, to the 
$\nu_e,\ \bar\nu_e$ and $\nu_{\mu,\tau}$ cross sections with electrons).}
\label{sigma}
\end{figure}

At $E_\nu\gsim 50$~MeV, 
the nucleon no longer looks like a point-like object
for the neutrinos, and hence the vector ($v_\mu$) and axial ($a_\mu$) 
hadronic currents
involve now momentum dependent form factors, i.e.
$$\langle N(p')|v_\mu|N(p)\rangle= \bar u(p')[
\gamma_\mu F_V+{i\over 2m_N}\sigma_{\mu\nu}q^\nu F_W] u(p)$$
$$\langle N(p')|a_\mu|N(p)\rangle= \bar u(p')[
\gamma_\mu \gamma_5F_A+{\gamma_5\over 2m_N}q_\mu F_P] u(p),$$
where $F_V(q^2)$ can be measured using electromagnetic processes 
and the CVC relation $F_V=F_V^{em,p}-F_V^{em,n}$ (i.e. as the difference
 between the proton and neutron electromagnetic 
vector form factors). Clearly $F_V(0)=1$ 
and $F_A(0)=1.26$, while $F_W$ is related to the magnetic moments
of the nucleons. The $q^2$ dependence has the effect of significantly 
flattening the cross section.
In the deep inelastic regime, $E_\nu\gsim$GeV, 
the neutrinos interact directly with the quark constituents. 
The cross section in this regime grows linearly with energy, and
this provided an important test of the parton model. 
The main characteristics of the neutrino cross section just discussed
are depicted in figure~2.

The final test of the standard model came with the direct production
of the $W^\pm$ and $Z$ gauge bosons at CERN in 1984, and with the 
precision test achieved with the $Z$ factories LEP and SLC after 1989.
 These
$e^+e^-$ colliders working at and around the $Z$ resonance ($s=M_Z^2 =
($91~GeV$)^2$) turned out to be also crucial for neutrino physics,
since  studying the shape of the $e^+e^-\to f\bar f$ cross section
near the resonance, which has the Breit--Wigner form
$$\sigma\simeq{12\pi \Gamma_e\Gamma_f\over M_Z^2} {s\over (s-M_Z^2)^2+
M_Z^2\Gamma_Z^2},$$
it becomes possible to determine the total $Z$ width $\Gamma_Z$. This
width is just the sum of all possible partial widths, i.e.
$$\Gamma_Z=\sum_f\Gamma_{Z\to f\bar f}=\Gamma_{vis}+\Gamma_{inv}.$$
The visible (i.e. involving charged leptons and quarks) width
$\Gamma_{vis}$ can be measured directly, and hence one can infer a value 
for the invisible width $\Gamma_{inv}$. Since in the standard model
this last arises from the decays $Z\to \nu_i\bar\nu_i$, whose 
expected rate for decays into a given neutrino flavour is
$\Gamma_{Z\to \nu\bar\nu}^{th}=167$~MeV, one can finally 
obtain the number of neutrinos coupling to the $Z$ as $N_\nu=\Gamma_{inv}/
\Gamma_{Z\to \nu\bar\nu}^{th}$. The present best value for this quantity
is $N_\nu=2.994\pm 0.012$, giving then a 
strong support to the three generation standard model.

\bigskip

Going through the history of the neutrinos we have seen that
they have been extremely useful to understand the standard model.
On the contrary, the standard model is
of little help to understand the neutrinos. Since in the 
standard model there is no need
for $\nu_R$, neutrinos are  massless in this theory. There is however 
no deep principle behind this (unlike the masslessness of the photon
which is protected by the electromagnetic gauge symmetry), and indeed 
in many extensions of the standard model neutrinos turn out to be 
 massive. This 
makes the search for non-zero neutrino masses a very important
issue, since it provides a window to look for physics beyond the
standard model.
There are many other important questions concerning the neutrinos 
which are  not addressed by the standard model, 
such as whether they are Dirac or Majorana particles, whether
lepton number is conserved, if the neutrino flavours are mixed (like
the quarks through the Cabibbo Kobayashi Maskawa matrix) and hence 
oscillate when they propagate, as many hints suggest today, whether they
have magnetic moments, if they decay, if they violate CP, and so on.
In conclusion, although in the standard model neutrinos are a little
bit boring,  many of its 
extensions contemplate new possibilities which 
make the neutrino physics a very exciting field.

\section{Neutrino masses:}

\subsection{Dirac or Majorana?}

In the standard model, charged leptons (and also quarks) 
get their masses through their
Yukawa couplings to the Higgs doublet field $\phi^T=(\phi_0,\phi_-)$
$$-{\cal L}_Y=\lambda \bar L\phi^*\ell_R+h.c.\ ,$$
where $L^T=(\nu,\ell)_L$ is a lepton doublet and $\ell_R$ an SU(2)
singlet field. 
When the electroweak symmetry gets broken by the vacuum expectation 
value of the neutral component of the Higgs field $\langle \phi_0\rangle
=v/\sqrt{2}$ (with $v=246$~GeV), the following `Dirac' mass term results
$$-{\cal L}_m=m_\ell (\bar\ell_L\ell_R+\bar\ell_R\ell_L)=m_\ell\bar\ell
\ell,$$
where $m_\ell=\lambda v/\sqrt{2}$ and $\ell=\ell_L+\ell_R$ is the Dirac 
spinor field. This mass term is clearly invariant under the $U(1)$
transformation $\ell\to {\rm exp}(i\alpha)\ell$, which 
corresponds to the lepton number (and actually in this case also
to the electromagnetic gauge invariance). From 
the observed fermion masses, one concludes that the Yukawa couplings 
range from $\lambda_t\simeq 1$ for the top quark up to $\lambda_e
\simeq 10^{-5}$ for the electron.

Notice that the mass terms always couple fields with 
opposite chiralities, i.e. requires a $L\leftrightarrow R$ transition.
Since in the standard model the right handed neutrinos are not 
introduced, it is not possible to write a Dirac mass term, and hence
the neutrino results massless. Clearly the simplest way to give the
neutrino a mass would be to introduce the right handed fields just
for this purpose (having no gauge interactions, these sterile states 
would be essentially undetectable and unproduceable). 
Although this is a logical possibility, it has the 
ugly feature that in order to get reasonable neutrino masses, below the 
eV, would require unnaturally small Yukawa couplings ($\lambda_\nu 
<10^{-11}$). Fortunately it turns out that neutrinos are also very
special particles in that, being neutral, there are other ways 
to provide them a mass. Furthermore, in some scenarios
 it becomes also possible to get a natural
understanding of why neutrino masses are so much smaller than the charged
fermion masses.

The new idea is that the left handed neutrino field actually involves
two degrees of freedom, the left handed neutrino associated with the
positive beta decay (i.e. emitted in association with a positron) and
the other one being  the right handed
`anti'-neutrino emitted in the negative beta decays 
(i.e. emitted in association with an electron). 
It may then be possible to write down
a mass term using just these two degrees of freedom and involving the 
required $L\leftrightarrow R$ transition. This possibility was first 
suggested by Majorana in 1937, in a paper named `Symmetric theory of 
the electron and positron', and devoted mainly to the problem of 
getting rid of the negative energy sea of the Dirac
equation\cite{ma37}.  As
a side product, he found that for neutral particles there was `no more
any reason to presume the existence of antiparticles', and that `it was 
possible to modify the theory of beta emission, both positive and 
negative, so that it came always associated with the emission of a 
neutrino'. The spinor field associated to this formalism was then
named in his honor a Majorana spinor.

To see how this works it is necessary to introduce the so called 
antiparticle field, $\psi^c\equiv C\bar\psi^T=C\gamma_0^T\psi^*$. 
The charge  conjugation matrix $C$ has to satisfy $C\gamma_\mu C^{-1}=
-\gamma_\mu^T$, so that for instance the Dirac equation for a 
charged fermion in the presence of an electromagnetic field, $(i\dslash 
-e\Aslash-m)\psi=0$ implies that  $(i\dslash 
+e\Aslash-m)\psi^c=0$, i.e. that the antiparticle field has opposite 
charges as the particle field and the same mass.
Since for a chiral projection one can show that $(\psi_L)^c=(P_L\psi)^c
=P_R\psi^c= (\psi^c)_R$, i.e. this conjugation changes the chirality of
the field, one has that $\psi^c$ is related to the $CP$ conjugate
of $\psi$. Notice  that $(\psi_L)^c$ describes exactly the same 
two degrees of freedom described by $\psi_L$, but somehow using a 
$CP$ reflected formalism. For instance for the neutrinos, the $\nu_L$ 
operator annihilates the left handed neutrino and creates the right
handed antineutrino, while the $(\nu_L)^c$ 
operator annihilates the right handed antineutrino and creates the left
handed neutrino.

We can then now write the advertised Majorana mass term, as
$$-{\cal L}_M={1\over 2}m[\overline{\nu_L}
(\nu_L)^c+\overline{(\nu_L)^c}\nu_L].$$
This mass term has the required Lorentz structure (i.e. the 
$L\leftrightarrow R$ transition) but one can see that it does not
preserve any $U(1)$ phase symmetry, i.e. it violates the so 
called lepton number by two units. 
If we introduce the Majorana field $\nu\equiv \nu_L+
(\nu_L)^c$, which under conjugation transforms into itself 
($\nu^c=\nu$), the mass term becomes just 
${\cal L}_M=-m\bar\nu\nu/2$.

\begin{figure}[t]
\begin{center}
\includegraphics[width=.2\textwidth,angle=-90]{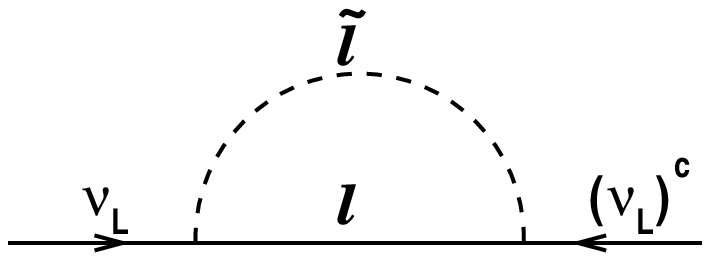}
\end{center} 
\caption{Example of loop diagram leading to a Majorana mass term
in supersymmetric models with broken $R$ parity.}
\label{rparity}
\end{figure}

Up to now we have introduced the Majorana mass by hand, 
contrary to the case of the charged fermions where it arose from a
Yukawa coupling in a spontaneously broken theory. To follow the same
procedure with the neutrinos presents however 
a difficulty,  
because the standard model neutrinos belong to $SU(2)$ doublets,
and hence to write an electroweak singlet Yukawa coupling 
it is necessary to introduce an $SU(2)$ triplet  Higgs field
$\vec\Delta$ (something which is not particularly attractive). 
The coupling  ${\cal L}\propto \overline{L^c} \vec\sigma L\cdot 
\vec\Delta$ would then lead to the Majorana mass term after the 
neutral component of the scalar gets a VEV. 
 Alternatively,
the Majorana mass term could be a loop effect in models where the
neutrinos have lepton number violating couplings to new scalars, 
as in the so-called Zee models or in the supersymmetric models
with $R$ parity violation (as illustrated in figure~\ref{rparity}).
These models have as interesting features that the masses are 
naturally suppressed by the loop, 
and they are attractive also if one looks 
for scenarios where the neutrinos have relatively large dipole 
moments, since a photon can be attached to the charged 
particles in the  loop.

\begin{figure}[t]
\begin{center}
\includegraphics[width=.3\textwidth,angle=-90]{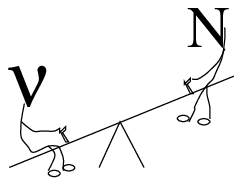}
\end{center} 
\caption{The see-saw model: pushing one mass up pushes the other down.}
\label{seesaw}
\end{figure}

However, by far the nicest possibility to give neutrinos a mass is the
so-called see-saw model introduced by Gell Man, Ramond and Slansky and by 
Yanagida in 1979\cite{sees}. 
In this scenario, which naturally occurs in grand unified
models such as $SO(10)$, one introduces the $SU(2)$ singlet right
handed neutrinos. One has now not only the ordinary Dirac mass term,
but also a Majorana
mass for the singlets which is generated by the VEV of an $SU(2)$ singlet
Higgs, whose natural scale is the scale of breaking of the grand
unified group, i.e. in the range $10^{12}$--$10^{16}$~GeV. Hence the 
Lagrangian will contain
$${\cal L}_M={1\over 2}\overline{(\nu_L,(N_R)^c)}\pmatrix{0&m_D\cr
m_D&M}\pmatrix{(\nu_L)^c\cr N_R}+ h.c..
\label{seesaweq}$$
The mass eigenstates are two Majorana fields with masses $m_{light}
\simeq m_D^2/M$ and $m_{heavy}\simeq M$. Since $m_D/M\ll 1$, we 
see that $m_{light}\ll
m_D$, and hence the lightness of the known neutrinos  is here 
related to the 
heaviness of the sterile states $N_R$, as figure~\ref{seesaw} 
illustrates.

If we actually introduce one singlet neutrino per family, the
mass terms in eq.~(\ref{seesaweq}) are $3\times 3$ matrices. Notice 
that if $m_D$ is similar to the up-type quark masses, as happens in 
$SO(10)$, one would have $m_{\nu_\tau}\sim m_t^2/M\simeq 
4$~eV$(10^{13}$~GeV/$M$).  It is clear then that in these scenarios
the observation of neutrino masses below the eV would point out to 
new physics at about the GUT scale.

\subsection{The quest for the neutrino mass:}
\subsubsection{Direct searches}

Already in his original paper on the theory of weak interactions Fermi
had noticed that the observed shape of the electron spectrum was
suggesting a small mass for the neutrino. The sensitivity to
$m_{\nu_e}$ in the decay $X\to X'+e+\bar\nu_e$ arises clearly because the
larger $m_\nu$, the less available kinetic energy remains for the
decay products, and hence the maximum electron energy is reduced. To
see this consider the phase space factor of the decay, $d\Gamma\propto
d^3 p_ed^3p_\nu\propto p_eE_edE_ep_\nu E_\nu dE_\nu \delta(E_e+E_\nu
-Q)$, with the $Q$--value being the total available energy in the
decay: $Q\simeq M_X-M_{X'}-m_e$. This leads to a differential electron
spectrum proportional to $d\Gamma /dE_e\propto p_eE_e(Q-E_e)
\sqrt{(Q-E_e)^2-m_\nu^2}$, whose shape near the endpoint ($E_e\simeq
Q-m_\nu$) 
depends on $m_\nu$ (actually the slope becomes infinite at the
endpoint for $m_\nu\neq 0$, while it vanishes for $m_\nu=0$).

Since the fraction of events in an interval $\Delta E_e$ around the
endpoint is $\sim (\Delta E_e/Q)^3$, to enhance the sensitivity to the
neutrino mass it is better to use processes with small $Q$-values,
what makes the tritium the most sensitive nucleus
($Q=18.6$~keV). Recent experiments at Mainz and Troitsk have allowed
to set the bound $m_{\nu_e}\leq 3$~eV. To improve this bound is quite
hard because the fraction of events within say 10~eV of the endpoint
is already $\sim 10^{-10}$.

Regarding the muon neutrino, a direct bound on its mass can be set by
looking to its effects on the available energy for the muon in the
decay of a pion at rest, $\pi^+\to\mu^++\nu_\mu$. From the knowledge
of the $\pi$ and $\mu$ masses, and measuring the momentum of the
monochromatic muon, one can get the neutrino mass through the relation 
$$m^2_{\nu_\mu}=m^2_\pi +m^2_\mu-2m_\pi\sqrt{p^2_\mu+m^2_\mu}.$$
The best bounds at present are $m_{\nu_\mu}\leq 170$~keV from PSI, and
again they are difficult to improve through this process since the
neutrino mass comes from the difference of two large quantities. There
is however a proposal to use  the muon $(g-2)$ experiment at BNL to
become sensitive down to $m_{\nu_\mu}\leq 8$~keV.

Finally, the bound on the $\nu_\tau$ mass is $m_{\nu_\tau}\leq 17$~MeV
and comes from the effects it has on the available phase space of the
pions in the decay $\tau\to 5\pi+\nu_\tau$ measured at LEP.

\begin{figure}[t]
\begin{center}
\includegraphics[width=1\textwidth]{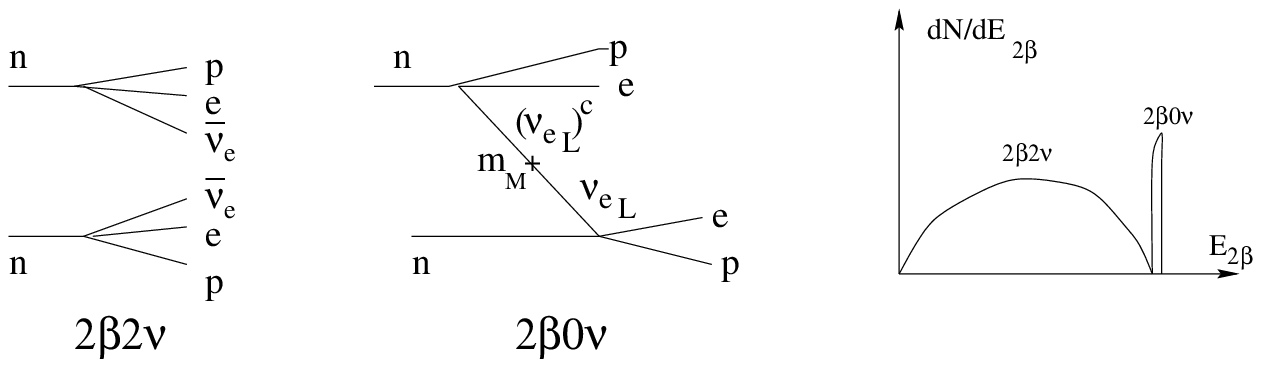}
\end{center} 
\caption{Double beta decay with and without neutrino emission, and
qualitative shape of the expected added spectrum of the two electrons.} 
\label{2beta}
\end{figure}

To look for the electron neutrino mass, besides the endpoint of the
ordinary beta decay there is another interesting process, but which is
however only sensitive to a Majorana (lepton number violating)
mass. This is the so called double beta decay. 
Some nuclei can undergo transitions in which two beta decays take
place simultaneously, with the emission of two electrons and two
antineutrinos ($2\beta 2\nu$ in fig.~\ref{2beta}). These transitions
have been observed in a few isotopes ($^{82}$Se, $^{76}$Ge,
$^{100}$Mo, $^{116}$Cd, $^{150}$Nd) in which the single beta decay is
forbidden, and the associated lifetimes are huge
($10^{19}$--$10^{24}$~yr). However, if the neutrino were a Majorana
particle, the virtual antineutrino emitted in one vertex could flip
chirality by a mass insertion and be absorbed in the second vertex as
a neutrino, as exemplified in fig.~\ref{2beta} ($2\beta 0\nu$). In
this way only two electrons would be emitted and this could be
observed as a monochromatic line in the added spectrum of the two
electrons. The non observation of this effect has allowed to set the
bound $m_{\nu_e}^{Maj}\leq 0.3$~eV (by the Heidelberg--Moscow
collaboration at Gran Sasso). There are projects to improve the
sensitivity of $2\beta 0\nu$ down to $m_{\nu_e}\sim 10^{-2}$~eV, and
we note that this bound is quite relevant since as we have seen, if
neutrinos are indeed massive it is somehow theoretically favored
(e.g. in the see saw models) that they are Majorana particles.

At this point it is important to extend the discussion to take into
account that there are three generations of neutrinos. If neutrinos
turn out to be massive, there is no reason to expect that the mass
eigenstates ($\nu_k$, with $k=1,2,3$) 
would coincide with the flavour (gauge) eigenstates ($\nu_\alpha$,
with $\alpha=e, \mu, \tau$), and
hence, in the same way that quark states are mixed through the
Cabibbo, Kobayashi and Maskawa matrix, neutrinos would be related
through the Maki, Nakagawa and Sakita
mixing matrix \cite{ma62}, i.e. $\nu_\alpha=V_{\alpha k}\nu_k$. The
MNS matrix  can be parametrized as ($c_{12}\equiv \cos\theta_{12}$,
etc.) 
\begin{eqnarray*}
V=\pmatrix{c_{12}c_{13} & c_{13}s_{12} & s_{13}\cr
-c_{23}s_{12}e^{i\delta}-c_{12}s_{13}s_{23} & c_{12}c_{23}e^{i\delta}
- s_{12}s_{13}s_{23} & c_{13}s_{23}\cr
s_{23}s_{12}e^{i\delta}-c_{12}c_{23}s_{13} & -c_{12}
s_{23}e^{i\delta} -c_{23}s_{12}s_{13} & c_{13}c_{23}}
\pmatrix{e^{i\alpha} & 0 & 0 \cr
0 & e^{i\beta} & 0 \cr
0 & 0 & 1}
\end{eqnarray*}
When the electron neutrino is a mixture of mass eigenstates, the
$2\beta 0\nu$ decay amplitude will be proportional now to an `effective
electron neutrino mass' $\langle m_{\nu_e}\rangle=V_{ek}^2m_k$, where
here we adopted the Majorana neutrino fields as self-conjugates
($\chi_k^c=\chi_k$). If one allows for Majorana creation phases
in the fields, $\chi_k^c=e^{i\alpha_k}\chi_k$, these phases will
appear in the effective mass,  $\langle
m_{\nu_e}\rangle={V'_{ek}}^2e^{i\alpha_k} m_k$. Clearly $\langle
m_{\nu_e}\rangle$ has to be independent of the unphysical phases
$\alpha_k$, so that the matrix diagonalising the mass matrix in the
new basis has to change accordingly,
i. e. $V'_{ek}=e^{-i\alpha_k/2}V_{ek}$. In particular, $\alpha$ and
$\beta$ may be removed from $V$ in this way, but they would anyhow
reappear at the end in $\langle m\rangle$ through the propagators of the
Majorana fields, which depend on the creation phases. 
When CP is conserved, it is
sometimes considered convenient to choose basis so that $V_{ek}$ is
real (i.e. $\delta=0$ from CP conservation and $\alpha$ and $\beta$
are reabsorbed in the Majorana creation phases of the fields). In this
case each contribution to $\langle m\rangle$ turns out to be 
 multiplied by the
intrinsic CP-parity of the mass eigenstate,  
$\langle
m_{\nu_e}\rangle=|\sum_k |V_{ek}|^2\eta_{CP}(\chi_k) m_k|$, with
$\eta_{CP}=\pm i$. States with opposite CP parities can then induce
cancellations in $2\beta 0\nu$ decays\footnote{In particular, Dirac neutrinos
can be thought of as two degenerate Majorana neutrinos with opposite
CP parities, and hence lead to a vanishing contribution to $2\beta
0\nu$, as would be expected from the conservation of lepton number in
this case.}.

Double beta decay is the only process sensitive to the phases $\alpha$
and $\beta$. These phases can be just phased away for Dirac neutrinos,
and hence in all experiments (such as oscillations) where it is not
possible to distinguish between Majorana and Dirac neutrinos, it is not 
possible to measure them. However, oscillation experiments are the
most sensitive way to measure small neutrino masses and their mixing
angles, as we now turn to discuss\footnote{Oscillations may even allow
to measure the CP violating phase $\delta$, e.g. by comparing
$\nu_\mu\to\nu_e$ amplitudes with the $\bar\nu_\mu\to\bar\nu_e$ ones,
as is now being considered for future neutrino factories at muon
colliders.}. 

\subsection{Neutrino oscillations:}

The possibility that neutrino flavour eigenstates be a superposition
of mass eigenstates, as was just  discussed,  allows for the
phenomenon of neutrino oscillations.  This is a quantum mechanical
interference effect (and as such it is sensitive to quite small masses)
and arises because different mass eigenstates propagate differently,
and hence the flavor composition of a state can change with time. 

\begin{figure}
\begin{center}
\includegraphics[width=1.3\textwidth,angle=-90]{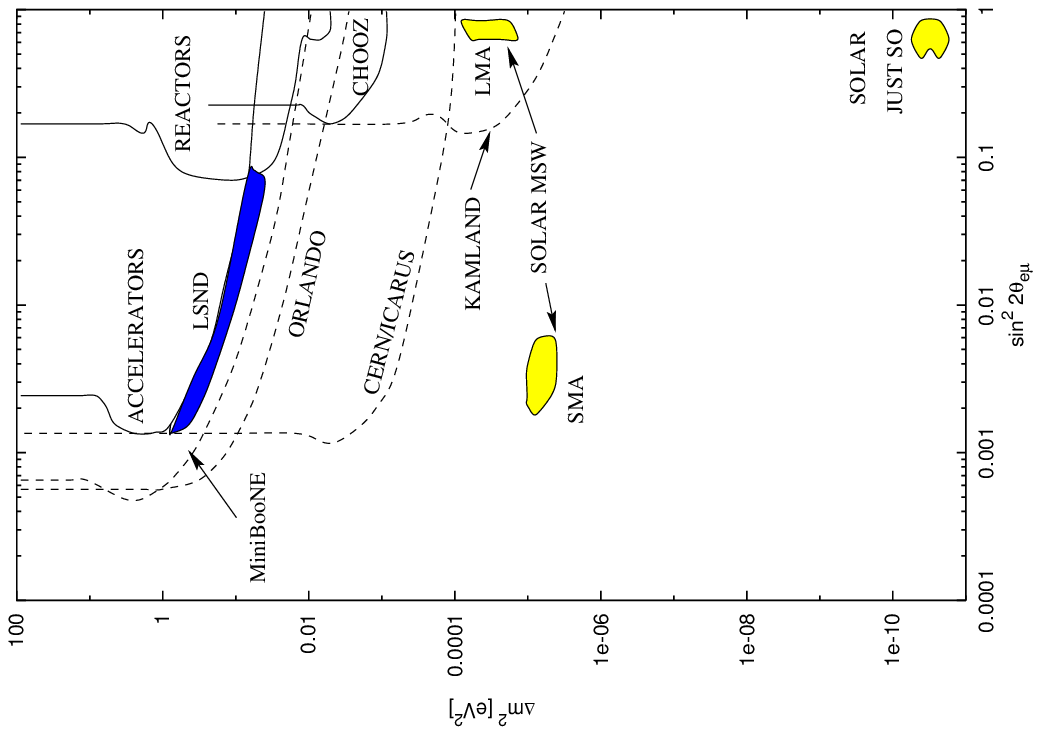}
\end{center} 
\caption{Present bounds (solid lines), projected sensitivities of
future experiments (dashed lines) and values suggested by LSND and
solar neutrino experiments for $\nu_e\leftrightarrow \nu_\mu$
oscillations.} 
\label{boundemu}
\end{figure}

To see this consider a flavour eigenstate neutrino $\nu_\alpha$ with
momentum $p$ produced at time $t=0$ (e.g. a $\nu_\mu$ produced in the
decay $\pi^+\to \mu^++\nu_\mu$). The initial state is then
$$|\nu_\alpha\rangle =\sum_kV_{\alpha k}|\nu_k\rangle .$$
We know that the mass eigenstates evolve with time according to 
 $|\nu_k(t,x) \rangle={\rm exp}[i(px-E_kt)]|\nu_k\rangle$. In the
relativistic limit relevant for neutrinos, one has that
$E_k=\sqrt{p^2+m_k^2}\simeq p+m^2_k/2E$, and thus the different mass
eigenstates will acquire different phases as they propagate.  Hence,
the probability of observing a flavour $\nu_\beta$ at time $t$ is just
$$ P(\nu_\alpha\to\nu_\beta)=|\langle \nu_\beta|\nu(t)\rangle|^2=
|\sum_kV_{\alpha k}e^{-i{m_i^2\over 2E}t}V^*_{\beta k}|^2.$$
In the case of two generations, taking $V$ just as a rotation with
mixing angle $\theta$, one has
$$ P(\nu_\alpha\to\nu_\beta)=\sin^22\theta\ \sin^2\left( {\Delta
m^2x\over 4E}\right),$$
which depends on the squared mass difference $\Delta
m^2=m_2^2-m_1^2$, since this is what gives the phase difference in the
propagation of the mass eigenstates. Hence, the amplitude of the
flavour oscillations is given by $\sin ^22\theta$ and the oscillation
length of the modulation is $L_{osc}\equiv {4\pi E\over \Delta
m^2}\simeq 2.5$~m $E$[MeV]/$\Delta m^2$[eV$^2$]. We see then that
neutrinos will typically oscillate with a macroscopic wavelength. For
instance, putting a detector at $\sim 100$~m from a reactor allows to
test oscillations of $\nu_e$'s to another flavour (or into a singlet
neutrino) down to $\Delta m^2\sim 10^{-2}$~eV$^2$ if sin$^22\theta$
is not too small ($\geq 0.1$). The CHOOZ experiment has even reached
$\Delta m^2\sim 10^{-3}$~eV$^2$ putting a large detector at 1~km
distance, and the proposed KAMLAND experiment will be sensitive to
reactor neutrinos arriving from $\sim 10^2$~km, and hence will test
$\Delta m^2\sim 10^{-5}$~eV$^2$ in a few years (see
fig.~\ref{boundemu}).

\begin{figure}
\begin{center}
\includegraphics[width=1.3\textwidth,angle=-90]{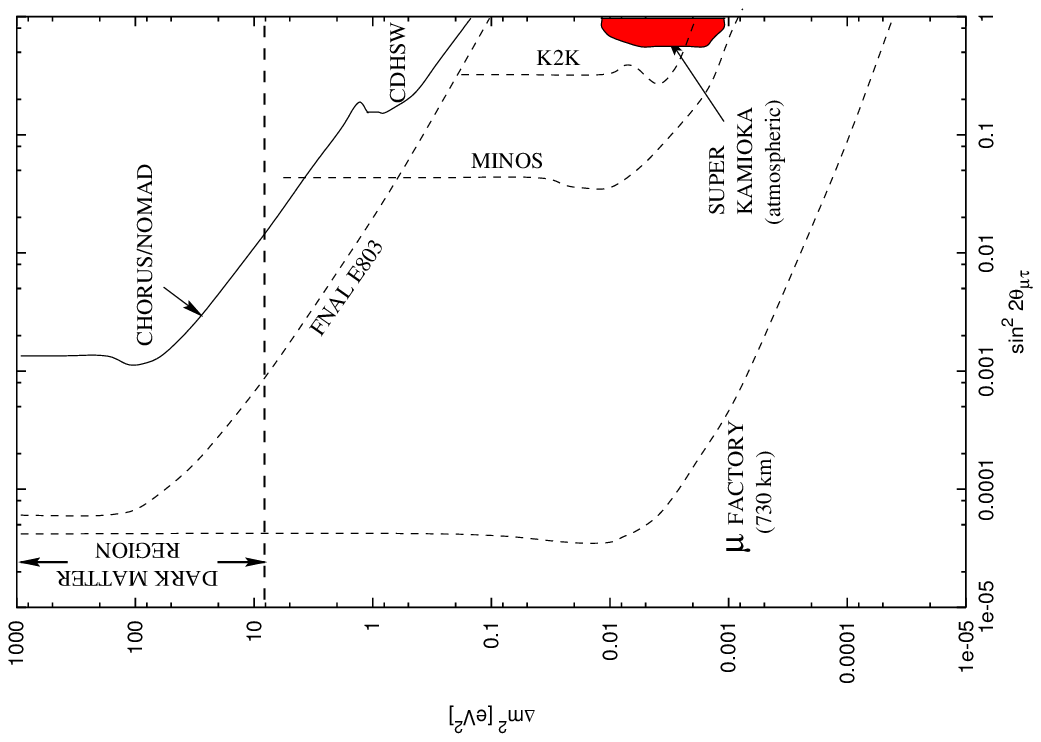}
\end{center} 
\caption{Present bounds (solid lines), projected sensitivities of
future experiments (dashed lines) and values suggested by the
atmospheric neutrino anomaly for  $\nu_\mu\leftrightarrow \nu_\tau$
oscillations. Also shown is the region where neutrinos would
constitute a significant fraction of the dark matter
($\Omega_\nu>0.1$).} 
\label{boundmutau}
\end{figure}

These kind of experiments look essentially for the disappearance of
the reactor $\nu_e$'s, i.e. to a reduction in the original $\nu_e$
flux. When one uses more energetic neutrinos from accelerators, it
becomes possible also to study the appearance of a flavour different
from the original one, with the advantage that one becomes sensitive
to very small oscillation amplitudes (i.e. small sin$^22\theta$
values), since the observation of only a few events is enough to
establish a positive signal. At present there is one experiment (LSND)
claiming a positive signal of $\nu_\mu\to\nu_e$ conversion, suggesting
the neutrino parameters in the region indicated in
fig.~\ref{boundemu}, once the region excluded by other experiments is
taken into account. The appearance of $\nu_\tau$'s out of a $\nu_\mu$
beam was searched at CHORUS and NOMAD at CERN without success,
allowing to exclude the region indicated in fig.~\ref{boundmutau},
which is a region of relevance for cosmology since neutrinos heavier
than $\sim$~eV would contribute to the dark matter in the Universe
significantly. 

In figs.~\ref{boundemu} and \ref{boundmutau} we also display the
sensitivity of various new experiments under construction or still at
the proposal level, showing that significant improvements are to be
expected in the near future (a useful web page with links to the
experiments is the Neutrino Industry Homepage\footnote{ 
http://www.hep.anl.gov/NDK/Hypertext/nuindustry.html}). These new
experiments  will in particular allow to test some of the most clear
hints we have at present in favor of massive neutrinos, which come
from the two most important natural sources of neutrinos that we have:
the atmospheric and the solar neutrinos.

\section{Neutrinos in astrophysics and cosmology:}

We have seen that neutrinos made their shy appearance in physics just
by steeling a little bit of the momentum of the electrons in a beta
decay. In astrophysics however, neutrinos have a major (sometimes
preponderant) role, being produced copiously in several environments.

\subsection{Atmospheric neutrinos:}

When a cosmic ray  (proton or nucleus) hits the atmosphere and knocks a
nucleus a few tens of km above ground, an hadronic (and
electromagnetic) shower is initiated, in which pions in particular are
copiously produced. The charged pion decays are the main source of
atmospheric neutrinos through the chain $\pi\to \mu\nu_\mu\to
e\nu_e\nu_\mu\nu_\mu$. One expects then twice as many $\nu_\mu$'s than
$\nu_e$'s  (actually at very high energies, $E_\nu\gg$ GeV, the parent
muons may reach the ground and hence be stopped before decaying, so
that the expected ratio $R\equiv
(\nu_\mu+\bar\nu_\mu)/(\nu_e+\bar\nu_e)$ should be even larger than
two at high energies). However, the observation of the atmospheric
neutrinos by IMB, Kamioka, Soudan, MACRO and Super Kamiokande
indicates that there is a deficit of muon neutrinos, with
$R_{obs}/R_{th}\simeq 0.6$ below $E_\nu\sim $~GeV. More remarkably, at
multi--GeV energies (for which a neutrino
 oscillation length would increase)
the Super Kamiokande experiment observes a zenith angle dependence
indicating that neutrinos coming from above (with pathlengths $d\sim
20$~km) had not  enough time to oscillate, while those from below
($d\sim 13000$~km) have already oscillated. 
The most plausible explanation for
these effects is an oscillation $\nu_\mu\to\nu_\tau$ with maximal
mixing and $\Delta m^2\simeq$ few$\times 10^{-3}$~eV$^2$, as indicated
in fig.~\ref{boundmutau}.

\subsection{Solar neutrinos:}

The sun gets its energy from the fusion reactions taking place in its
interior, where essentially four protons form a He nucleus. By charge
conservation this has to be accompanied by the emission of two
positrons and by lepton number conservation in the weak processes two
$\nu_e$'s have to be produced. This fusion liberates 27~MeV of
energy, which is eventually emitted mainly (97\%) as photons and the
rest (3\%) as neutrinos. Knowing the energy flux of the solar
radiation reaching the Earth ($k_\odot\simeq 1.5$~kW/m$^2$), it is
then simple to estimate that the solar neutrino flux at Earth is
$\Phi_\nu\simeq 2k_\odot/27$~MeV $\simeq 6\times
10^{10}\nu_e/$cm$^2$s, which is a very large number indeed.

Many experiments  have looked for these solar neutrinos and the
puzzling result which has been with us for the last thirty years is
that only between 1/2 to 1/3 of the expected fluxes are
observed. Remarkably, Pontecorvo \cite{po67} noticed even before the
first observation of solar neutrinos by Davies that neutrino
oscillations could reduce the expected rates. We note that the
oscillation length of solar neutrinos ($E\sim 0.1$--10~MeV) is of the
order of 1 AU for $\Delta m^2\sim 10^{-10}$~eV$^2$, and hence even
those tiny neutrino masses can have observable effects if the mixing
angles are large (this would be the `just so' solution to the solar
neutrino problem). Much more remarkable is the possibility of
explaining the puzzle by resonantly enhanced oscillations of neutrinos
as they propagate outwards through the Sun. Indeed, the solar medium
affects $\nu_e$'s differently than $\nu_{\mu,\tau}$'s (since only the
first interact through charged currents with the electrons present),
and this modifies the oscillations in a beautiful way through an
interplay of neutrino mixings and matter effects, in the so called MSW
effect \cite{msw}. Two possible solutions using this mechanism require
$\Delta m^2\simeq 10^{-5}$~eV$^2$ and small mixings $s^22\theta\simeq$
few$\times 10^{-3}$~eV$^2$ (SMA) or large mixing (LMA), as shown in
fig.~\ref{boundemu}. 

Atmospheric and solar neutrinos are extremely fashionable
nowadays. For  instance more than a dozen review papers on the subject
have appeared in the last year and hence I will avoid going with more
details into them (see e.g. \cite{ge95,reviews}), although the second
lecture dealt exclusively with this subject.

\subsection{Supernova neutrinos:}

The most spectacular neutrino fireworks in the Universe are the
supernova explosions, which correspond to the death of a very massive
star. In this process the inner Fe core ($M_c\simeq 1.4\ M_\odot$),
unable to get pressure support gives up to the pull of gravity and
collapses down to nuclear densities (few$\times 10^{14}$g/cm$^3$),
forming a very dense proto-neutron star. At this moment neutrinos
become the main character on stage, and 99\% of the gravitational
binding energy gained (few$\times 10^{53}$~ergs) is released in a
violent burst of neutrinos and antineutrinos of the three flavours,
with typical energies of a few tens of MeV\footnote{Actually there is
first a brief (msec) $\nu_e$ burst from the neutronisation of the Fe
core.}. 
Being the density so high, even the weakly interacting neutrinos
become trapped in the core, and they diffuse out in a few seconds to be
emitted from the so called neutrinospheres (at $\rho\sim
10^{12}$~g/cm$^3$). These neutrino fluxes then last for $\sim 10$~s,
after which the initially trapped lepton number is lost and the
neutron star cools more slowly. 

During those $\sim 10$~s the neutrino luminosity of the supernova
($\sim 10^{52}$~erg/s) is comparable to the total luminosity of the
Universe (c.f. $L_\odot\simeq 4\times 10^{33}$~erg/s), but
unfortunately  only a couple of such events occur in our galaxy per
century, so that one has to be patient. Fortunately, on February 1987
a supernova exploded in the nearby ($d\sim 50$~kpc) 
Large Magellanic Cloud,
producing a dozen neutrino events in the Kamiokande and IMB
detectors. This started extra solar system neutrino astronomy and
provided a very basic proof of the explosion mechanism. With the new
larger detectors under operation at present (SuperKamioka and SNO) it
is expected that a future galactic supernova ($d\sim 10$~kpc) would
produce several thousand neutrino events and hence allow detailed
studies of the supernova physics. 

Also sensitive tests of neutrino properties will be feasible if a
galactic supernova is observed. The simplest example being the limits
on the neutrino mass which would result from the measured burst
duration as a function of the neutrino energy. Indeed, if neutrinos
are massive, their velocity will be $v=c\sqrt{1-(m_\nu/E)^2}$, and
hence the travel time  from a SN at distance $d$ would be $t\simeq 
{d\over
c}[1-{1\over 2}(m_\nu/E)^2]$, implying that lower energy neutrinos
($E\sim 10$~MeV) would arrive later than  high energy ones by an
amount $\Delta t\sim 0.5 (d/10$~kpc) ($m_\nu/10$~eV)$^2$s. Looking for
this effect
 a sensitivity down to $m_{\nu_{\mu,\tau}}\sim 25$~eV would be
achievable from a supernova at 10~kpc, and this is much better than
the present direct bounds on the $\nu_{\mu,\tau}$ masses.

\begin{figure}[t]
\begin{center}
\includegraphics[width=0.8\textwidth]{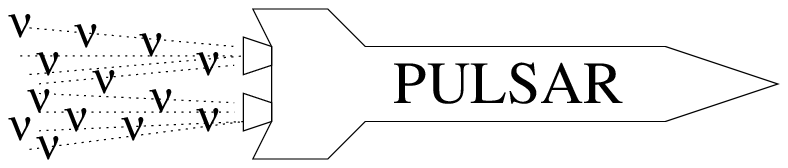}
\end{center} 
\caption{Pulsar kicks from neutrino rockets?}
\label{rocket}
\end{figure}

What remains after a (type II) supernova explosion is a pulsar, i.e. a
fastly rotating magnetised neutron star. One of the mysteries related
to pulsars is that they move much faster (few hundred km/s) than
their progenitors (few tenths of km/s). There is no satisfactory
standard explanation of how these initial kicks are imparted to the
pulsar and here neutrinos may also have something to say. It has been
suggested that these kicks could be due to a macroscopic manifestation
of the parity violation of weak interactions, i.e. that in the same
way as electrons preferred to be emitted in the direction opposite to
the polarisation of the $^{60}$Co nuclei in the experiment of Wu 
(and hence the neutrinos preferred to be emitted in the same
direction), the neutrinos in the supernova explosions would be biased
towards one side of the star because of the polarisation induced in
the matter by the large magnetic fields present \cite{ch84}, leading to
some kind of neutrino rocket effect, as shown in fig.~\ref{rocket}. 

Although only a 1\% asymmetry in the emission of the neutrinos would
be enough to explain the observed velocities, the magnetic fields
required are $\sim 10^{16}$~G,
much larger than the ones inferred from
observations ($\sim 10^{12}$--$10^{13}$~G).
An attempt has also been done \cite{ku96} 
to exploit the fact that the neutrino
oscillations in matter are affected by the magnetic field, and hence
the resonant flavor conversion would take place in an off-centered
surface. Since $\nu_\tau$'s (or $\nu_\mu$'s) interact less than
$\nu_e$'s, an oscillation from $\nu_e\to\nu_\tau$ in the region where
$\nu_e$'s are still trapped but $\nu_\tau$'s can freely escape would
generate a $\nu_\tau$ flux from a deeper region of the star in one
side than in the other. Hence if one assumes that the temperature
profile is isotropic, neutrinos from the deeper side will be more
energetic than those from the opposite side and can then be the source
of the kick. This would require however $\Delta m^2>(100$~eV)$^2$, which
is uncomfortably large, and $B>10^{14}$~G. Moreover, it has been
argued \cite{ja98} 
that the assumption of an isotropic $T$ profile near the
neutrinospheres will not hold, since the side where the escaping
neutrinos are more energetic will rapidly cool (the neutrinosphere
region has negligible heat capacity compared to the core) adjusting
the temperature gradient so that 
the isotropic energy flux generated in the core will manage ultimately
to get out isotropically.

An asymmetric neutrino emission due to an asymmetric magnetic field
affecting asymmetrically the $\nu_e$ opacities has also been proposed,
but again the magnetic fields required are too large ($B\sim
10^{16}$~G) \cite{opac}.

As a summary, to explain the pulsar kicks as due to an asymmetry in
the neutrino emission is attractive theoretically, but unfortunately
doesn't seem to work. Maybe when three dimensional simulations of the
explosion would become available, possibly including the presence of a
binary companion, larger asymmetries would be found just from standard 
hydrodynamical processes.

Supernovae are also helpful for us in that they throw away into the
interstellar medium all the heavy elements produced during the star's
life, which are then recycled into second generation stars like the
Sun, planetary systems and so on. However, 25\% of the baryonic mass
of the Universe was already in the form of He nuclei well before the
formation of the first stars, and as we understand now this He was
formed a few seconds after the big bang in the so-called primordial
nucleosynthesis. Remarkably, the production of this He also depends on
the neutrinos, and the interplay between neutrino physics and
primordial nucleosynthesis provided the first important astro-particle
connection. 

\subsection{Cosmic neutrino background and primordial nucleosynthesis:}

In the same way as the big bang left over the 2.7$^\circ$K cosmic
background radiation, which decoupled from matter after the
recombination epoch ($T\sim $~eV), there should also be a relic
background of cosmic neutrinos (C$\nu$B) left over from an earlier
epoch ($T\sim$~MeV), when weakly interacting neutrinos decoupled from
the $\nu_i$--$e$--$\gamma$ primordial soup. Slightly after the
neutrino decoupling, $e^+e^-$ pairs annihilated and reheated the
photons, so that the present temperature of the C$\nu$B is
$T_\nu\simeq 1.9^\circ$K, slightly smaller than the photon one. This
means that there should be today a density of neutrinos (and
antineutrinos) of each flavour $n_{\nu_i}\simeq 110$~cm$^{-3}$. 

Primordial nucleosynthesis occurs between $T\sim 1$~MeV and
$10^{-2}$~MeV, an epoch at which the density of the Universe was
dominated by radiation (photons and neutrinos). This means that the
expansion rate of the Universe depended on the number of neutrino
species $N_\nu$, becoming faster the bigger $N_\nu$
($H\propto\sqrt{\rho} \propto\sqrt{\rho_\gamma+N_\nu\rho_\nu}$, with
$\rho_\nu$ the density for one neutrino species). 
Helium production just occurred after deuterium
photodissociation became inefficient at $T\sim 0.1$~MeV, with
essentially all neutrons present at this time ending up into He. The
crucial point is that the faster the expansion rate, the larger
fraction of neutrons (w.r.t protons) would have survived to produce He
nuclei. This implies that an observational upper bound on the primordial
He abundance will translate into an upper bound on the number
of neutrino species. Actually the predictions also depend in the total
amount of baryons present in the Universe ($\eta=n_B/n_\gamma$), which
can be determined studying the very small amounts of primordial D and
$^7$Li produced. The observational D measurements are somewhat
unclear at presents, with determinations in the low side implying the
strong constraint $N_\nu<3.3$, while those in the high side implying
$N_\nu<4.8$ \cite{ol99}.
It is important that nucleosynthesis 
bounds on $N_\nu$ were established well before
the LEP measurement of the number of standard neutrinos.

As a side product of primordial nucleosynthesis theory one can
determine that the amount of baryonic matter in the Universe has to
satisfy $\eta\simeq 1$--$6\times 10^{-10}$. The explanation of this
small number is one of the big challenges for particle physics and
another remarkable fact of neutrinos is that they might be ultimately
responsible for this matter-antimatter asymmetry.

\subsection{Leptogenesis:}

\begin{figure}[t]
\begin{center}
\includegraphics[width=0.8\textwidth]{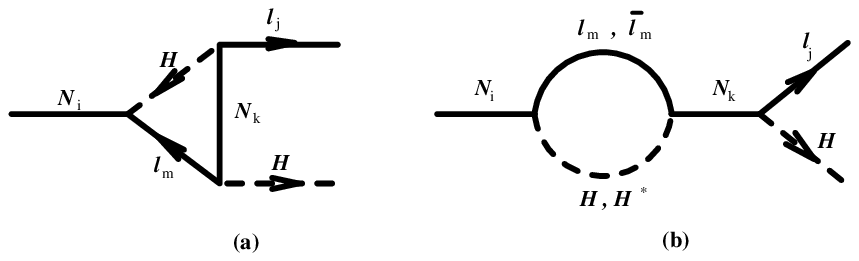}
\end{center} 
\caption{One loop CP violating diagrams for leptogenesis.}
\label{lepto}
\end{figure}

The explanation of the observed baryon asymmetry as due to
microphysical processes taking place in the early Universe is known to
be possible provided the three Sakharov conditions are fulfilled: $i)$
the existence of baryon number violating interactions ($\viola{B}$),
$ii)$ the existence of C and CP violation ($\viola{C}$ and
$\rlap{\it C}{\,\,\not P}$) 
and $iii)$ departure from chemical equilibrium ($\rlap{\it E}{\,\,\not q}$). The
simplest scenarios fulfilling these conditions appeared in the
seventies with the advent
of GUT theories, where heavy color triplet Higgs bosons 
can decay out of equilibrium
in the rapidly expanding Universe (at $T\sim M_T\sim 10^{13}$~GeV)
violating B, C and CP. In the middle of the eighties it was realized
however that in the Standard Model non-perturbative $\viola{B}$ and
$\viola{L}$ (but $B-L$ conserving) processes where in equilibrium at
high temperatures $(T>100$~GeV), and would lead to a transmutation
between $B$ and $L$ numbers, with the final outcome that $n_B\simeq
n_{B-L}/3$. This was a big problem for the simplest GUTs like SU(5),
where $B-L$ is conserved (and hence $n_{B-L}=0$), but it was rapidly
turned into a virtue by Fukugita and Yanagida  \cite{fu86}, who
realised that it could be sufficient to generate initially a lepton
number asymmetry and this  will then be reprocessed into a baryon number
asymmetry. The nice thing is that in see-saw models the generation of
a lepton asymmetry (leptogenesis) is quite natural, since the heavy
singlet Majorana neutrinos would decay out of equilibrium (at $T\lsim
M_N$) through $N_R\to\ell H,\ \bar\ell H^*$, i.e. into final states
with different $L$, and the CP violation appearing at one loop through
the diagrams in fig.~\ref{lepto} would lead \cite{co96} to
$\Gamma(N\to\ell H)\neq\Gamma(N\to\bar\ell H^*)$, so that a final $L$
asymmetry will result. Reasonable parameter values lead naturally to
the required asymmetries ($\eta\sim 10^{-10}$), making this scenario
probably the simplest baryogenesis mechanism.

\subsection{Neutrinos as dark matter:}

Neutrinos may not only give rise to the observed baryonic matter, but
they could also themselves be the dark matter in the Universe. This
possibility arises \cite{ge66} because if the ordinary neutrinos are
massive, the large number of them present in the C$\nu$B will
significantly contribute to the mass density of the Universe, in an
amount\footnote{The reduced Hubble constant is $h\equiv
H/(100$~km/s-Mpc$)\simeq 0.6$.} $\Omega_\nu\simeq
\sum_im_{\nu_i}/(92h^2$~eV). Hence, in order for neutrinos not to
overclose the Universe it is necessary that $\sum_im_{\nu_i}\lsim
30$~eV, which is a bound much stronger than the direct ones for
$m_{\nu_{\mu,\tau}}$. On the other hand, a neutrino mass $\sim 0.1$~eV
(as suggested by the atmospheric neutrino anomaly) would imply that
the mass density in neutrinos is already comparable to that in
ordinary baryonic matter ($\Omega_B\sim 0.003$), and $m_\nu\gsim 1$~eV
would lead to an important contribution of neutrinos to the dark
matter.

\begin{figure}[t]
\begin{center}
\includegraphics[width=.58\textwidth,angle=-90]{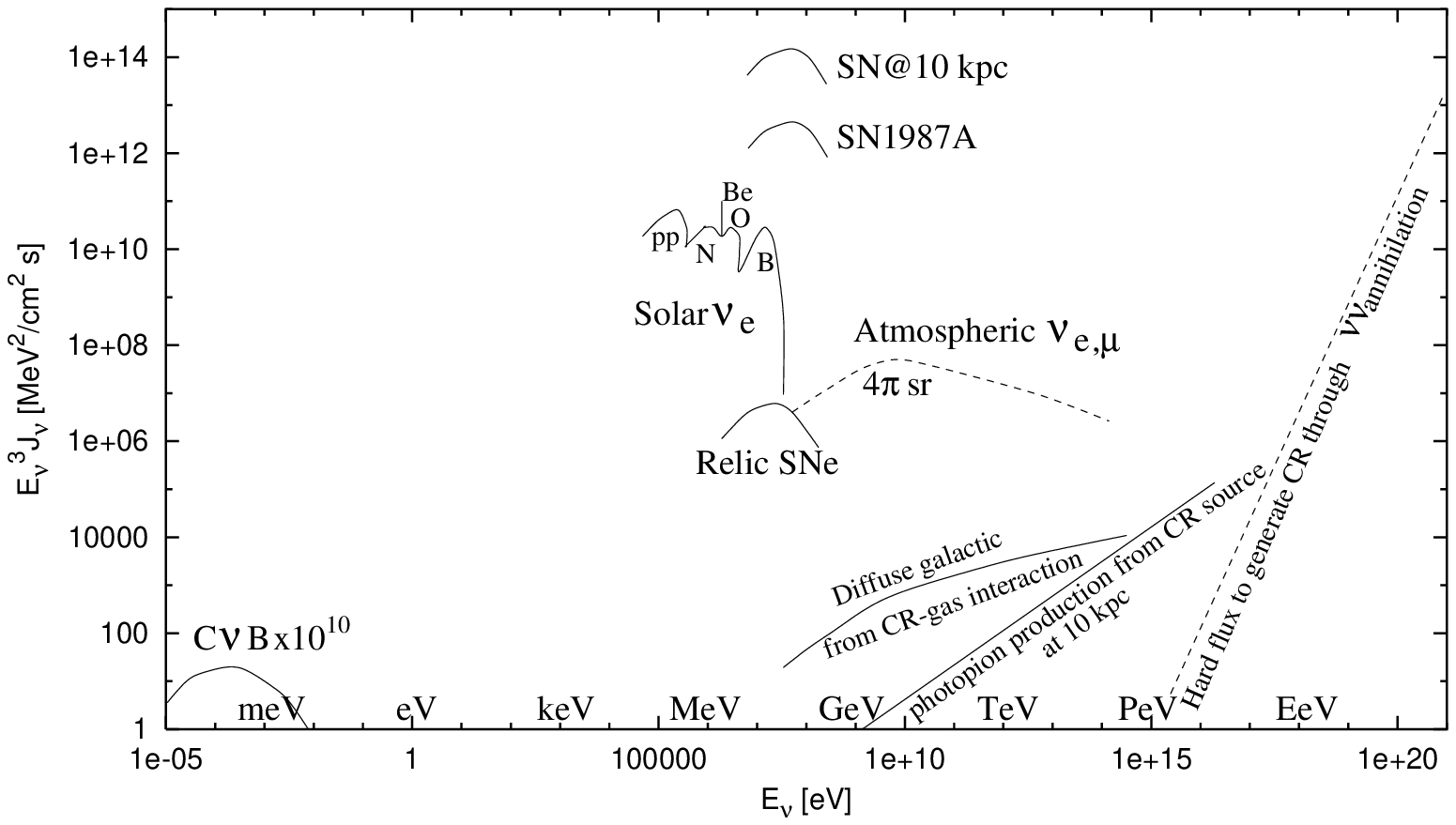}
\end{center} 
\caption{Neutrino spectra. We show the cosmic neutrino background
(C$\nu$B) multiplied by $10^{10}$, solar and supernova neutrinos, the
isotropic atmospheric neutrinos, those coming from the galactic plane
due to cosmic ray gas interactions, an hypothetical galactic source at
10~kpc, whose detection at $E>10$~TeV would require a good angular
resolution to reject the atmospheric background (similar
considerations hold for AGN neutrinos not displayed). 
Finally the required flux to
produce the CR beyond the GZK cutoff by annihilations with the dark
matter neutrinos.}
\label{spectra}
\end{figure}

The nice things of neutrinos as dark matter is that they are the only
candidates that we know for sure that they exist, and that they are
very helpful to generate the structures observed at large supercluster
scales ($\sim 100$~Mpc). However, they are unable to give rise to
structures at galactic scales (they are `hot' and hence free stream
out of small inhomogeneities).  Furthermore, even if those structures
were formed, it would not be possible to pack the neutrinos
sufficiently so as to account for the galactic dark halo densities,
due to the lack of sufficient phase space \cite{tr79}, since to
account for instance for the local halo density $\rho^0\simeq
0.3$~GeV/cm$^3$ would require $n_\nu^0\simeq
10^8(3$~eV$/m_\nu)$/cm$^3$, which is a very large overdensity with
respect to the average value 110/cm$^3$. The Tremaine Gunn phase-space
constraint requires for instance that to be able to account for the
dark matter in our galaxy one neutrino should be heavier than $\sim
50$~eV, so that neutrinos can clearly account at most for a fraction
of the galactic dark matter.

The direct detection of the dark matter neutrinos will be extremely
difficult \cite{la83}, 
because of their very small energies ($E\simeq m_\nu
v^2/2\simeq 10^{-6}m_\nu c^2$) which leads to very tiny cross sections
with matter and involving tiny momentum transfers. This has lead people to
talk about kton detectors at mK temperatures in zero gravity
environments ..., and hence this remains clearly as a challenge for
the next millennium.

One speculative proposal to observe the dark matter neutrinos
indirectly is through the observation  of the annihilation of cosmic
ray neutrinos of ultra high energies $(E_\nu\sim
10^{21}$~eV/$(m_\nu/4$~eV)) with dark matter ones at the $Z$-resonance
pole where the cross section is enhanced \cite{we82}. Moreover, this
process has been suggested as a possible way to generate the observed
hadronic cosmic rays above the GZK cutoff \cite{we99}, since neutrinos
can travel essentially unattenuated for cosmological distances ($\gg
100$~Mpc) and induce hadronic cosmic rays locally through their
annihilation with dark matter neutrinos. This proposal requires
however extremely powerful neutrino sources.

In fig.~\ref{spectra} we summarize qualitatively different fluxes
which can appear in the neutrino sky and whose search and observation
is opening new windows to understand the Universe.

\section*{Acknowledgments}

This work was supported by CONICET and Fundaci\'on Antorchas.

%

\end{document}